\documentclass[conference,a4paper,comsoc]{IEEEtran}
\usepackage[nocompress]{cite}
\usepackage{epsfig}
\usepackage{amsmath,graphicx,subfig}
\usepackage{tikz}
\usepackage{multicol}
\usepackage{multirow}
\usepackage{balance}
\usepackage{nicefrac}
\usetikzlibrary{decorations.pathreplacing}
\usetikzlibrary{chains,arrows,shapes,spy}
\usepackage{caption}
\usepackage{enumitem}
\usepackage{url}
\usetikzlibrary{calc}
\definecolor{r}{rgb}{0, 0, 0}
\definecolor{rc}{rgb}{0, 0, 0}
\definecolor{r}{rgb}{0, 0, 0}
\definecolor{c}{rgb}{0, 0, 0}
\tikzset{
	font={\fontsize{8.5}{11.0476pt}\selectfont}}
\usepackage{mathtools,amssymb,cuted}
\usepackage{pgfplots}
\pgfplotsset{compat=newest}
\usepackage{romannum}

\usepackage{algpseudocode}
\usepackage{algorithm}
\algnewcommand{\Parameters}[1]{%
	\State \textbf{\underline{Parameters}:}
	\Statex \hspace*{\algorithmicindent}\parbox[t]{.8\linewidth}{\raggedright #1}
}

\algnewcommand{\Input}[1]{%
	\State \textbf{\underline{Inputs}:}
	\Statex \hspace*{\algorithmicindent}\parbox[t]{.8\linewidth}{\raggedright #1}
}

\algnewcommand{\Initialization}[1]{%
	\State \textbf{\underline{Initialization}:}
	\Statex \hspace*{\algorithmicindent}\parbox[t]{.8\linewidth}{\raggedright #1}
}

\algnewcommand{\Iteration}[1]{%
	\State \textbf{\underline{Iteration}:}
	\Statex \hspace*{\algorithmicindent}\parbox[t]{.8\linewidth}{\raggedright #1}
}
\newcommand{\cev}[1]{\reflectbox{\ensuremath{\vec{\reflectbox{\ensuremath{#1}}}}}}
\def\BibTeX{{\rm B\kern-.05em{\sc i\kern-.025em b}\kern-.08em
    T\kern-.1667em\lower.7ex\hbox{E}\kern-.125emX}}

\begin{document}
\title{Study of Constrained Precoding with Zero-Crossing Modulation for Channels with 1-Bit Quantization and Oversampling}

\author{
    \IEEEauthorblockN{
        Diana M. V. Melo,
        ~Lukas T. N. Landau,
        ~and Rodrigo C. de Lamare,}
	\IEEEauthorblockA{
	    Centre for Telecommunications Studies, Pontifical Catholic University of Rio de Janeiro, \\
     Rio de Janeiro, Brazil 22453-900\\
	    Email: diana@aluno.puc-rio.br;landau;delamare@puc-rio.br
    }
    \vspace{-8mm}
}
\maketitle

\begin{abstract}
Future wireless communications systems are expected to operate at bands above 100GHz. The high energy consumption of analog-to-digital converters, due to their high resolution represents a bottleneck for future wireless communications systems which require low-energy consumption and low-complexity devices at the receiver. In this work, we derive a novel precoding method based on quality of service constraints for a multiuser multiple-input multiple-output downlink system with 1-bit quantization and oversampling. For this scenario, the time-instance zero-crossing modulation which conveys the information into the zero-crossings is considered. Unlike prior studies, the constraint is given regarding the symbol error probability related to the minimum distance to the decision threshold. Numerical results illustrate the performance of the proposed precoding method evaluated under different parameters.

\end{abstract}

\section{Introduction}

Future wireless communications systems are expected to operate at millimeter-wave and sub-terahertz frequency bands and support a massive number of devices \cite{Rappaport2019} as in Internet of Things (IoT) scenarios \cite{Gupta_2015} and reach higher data rates \cite{Viswanathan_2020}. The transmission of higher data rates can represent a challenge in terms of the design of energy-efficient analog
to-digital converters (ADCs) since the power consumption in the ADC increases exponentially with amplitude resolution \cite{Murmann_2009} and quadratically with the sampling rate for bandwidths above 300MHz \cite{Murmann_ADC,cqabd}.
An established approach to decreasing the power consumption of each ADC is to consider 1-bit quantization \cite{bbprec,1bitidd,dqalms,dqarls,dynovs,comp}. In addition to reducing energy consumption, it reduces the complexity of the devices since automatic gain control can potentially be omitted. The loss of amplitude information can be compensated by increasing the sampling rate \cite{Gilbert_1993}. In a noise-free case, it has been shown that rates of $\log_2(M_\mathrm{Rx}+1)$ bits per Nyquist interval are achievable by $M_\mathrm{Rx}$-fold oversampling with respect to (w.r.t.) the Nyquist rate \cite{Shamai2_1994}.

The information must be encoded into the temporal samples with one-bit resolution. {Methods based on 1-bit quantization and oversampling have been introduced in \cite{Landau_2014, Son_2019,1bitcpm,dynovs}}. In \cite{Shamai2_1994} the constructed bandlimited transmit signal conveys the information into zero-crossing patterns. In this sense modulation schemes based on zero-crossing have been proposed in \cite{Peter_2021, Peter_2020, Bender_2019} with runlength limited (RLL) transmit sequences and \cite{Viveros_2023} with the time instance zero-crossing (TI ZX) modulation. The TI ZX modulation from \cite{Viveros_2023} encodes the information into the time-instance of zero-crossings in order to reduce the number of zero-crossings of the signal. Important precoding approaches for MIMO channels have been studied for systems with 1-bit quantization modulation based on the maximization of the minimum distance to the decision threshold (MMDDT) \cite{Viveros_2023}, minimum mean squared error (MMSE) \cite{Viveros_2023, Amine_2016, Jacobson_2016}, state-machine based waveform design optimization \cite{Viveros_2024} and quality of service (QOS) constraint \cite{Viveros_ssp_2021}. The proposed method in \cite{Viveros_ssp_2021}  minimizes the transmit power while taking into account quality of service constraints in terms of the minimum distance to the decision threshold. Some of these methods improve performance when combined with faster-than-Nyquist (FTN) signaling \cite{Mazo_1975}. Moreover, in \cite{Viveros_2023} a spectral efficiency lower bound is presented for the TI ZX modulation. Besides, in \cite{Viveros_2023} an analytical method is introduced for the TI ZX modulation with MMSE precoding. {Moreover, the study in \cite{Erico_2023} proposes a branch-and-bound method with  QOS for a solution that attains a target symbol-error probability}. \textcolor{r}{Other modulation schemes have been presented in \cite{mohamed_2017} for multiple-input multiple-output (MIMO) systems.}

 In this study, we consider a bandlimited multiuser MIMO downlink system with 1-bit quantization and oversampling, considering the TI ZX modulation \cite{Viveros_2023}. For this study, a semi-analytical symbol error rate upper bound is presented for the QOS constraint method with an established minimum distance to the decision threshold \cite{Viveros_ssp_2021}. {Different from \cite{Viveros_ssp_2021}, the precoding method is formulated such that the constraint is given in terms of a target  SER. Moreover considering the Gray coding for TI ZX modulation from \cite{Viveros_2023}, an approximate BER can also be defined as constraint. Numerical results illustrate the performance of the proposed and other competing techniques.} 
 
 The rest of this paper is organized as follows, Section~\ref{sec:system_model} details the considered system model. 
Afterwards, in Section~\ref{sec:TIZX} we explain the TI ZX modulation and the detection process. Section~\ref{sec:QOS} presents the optimization problem for QOS temporal precoding method. Then, the performance bound is introduced in Section~\ref{sec:bound}. Numerical results are presented in Section~\ref{sec:num_results} and
finally, we conclude in Section~\ref{sec:conclusiones}.

Notation: In the paper, all scalar values, vectors, and matrices are represented by $a$, ${\boldsymbol{x}}$ and ${\boldsymbol{X}}$, respectively. \textcolor{c}{The subscript $I/Q$ means that the process is done separately for the signal's phase and quadrature component.}

\section{System Model}
\label{sec:system_model}

\begin{figure}[t]
\begin{center}
\captionsetup{justification=centering}
\includegraphics[height=5cm, width=8cm]{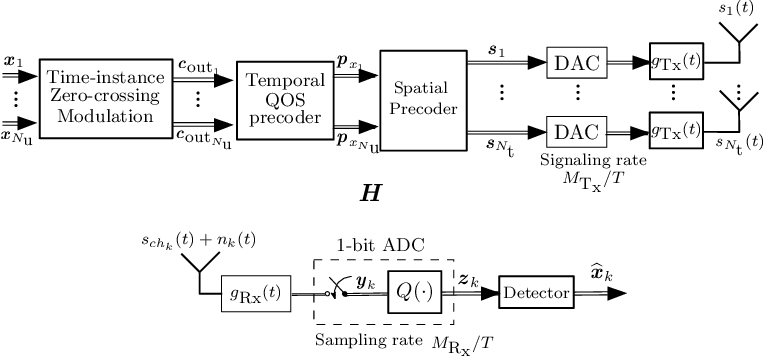}
\caption{Multiuser MIMO system model with $N_{\text{u}}$ single antenna users.}
\label{fig:system_mimomodel}       
\vspace{-0.8em}
\end{center}
\vspace{-0.6em}
\end{figure}

The multiuser MIMO system model \cite{mmimo,wence} is considered in the downlink with  $N_{\text{t}}$ transmit antennas and $N_{\text{u}}$ single antenna users as depicted in Fig.~\ref{fig:system_mimomodel}. The vector $\boldsymbol{x}_{k} \in \mathbb{C}$ for user $k$, with $N$ symbols corresponds to the transmit vector which is mapped into the TI ZX modulated pattern $\boldsymbol{c}_{\textrm{out}} \in \mathbb{C}^{N_{\textrm{tot}}\times 1}$ where $N_{\textrm{tot}} = N M_{\textrm{Rx}}+1$ and $M_{\textrm{Rx}}$ corresponds to the oversampling factor. Then, the space-time precoding vector $\boldsymbol{p}_{x} \in \mathbb{C}^{N_{\textrm{q}}{N_{\textrm{T}}}\times 1}$ is obtained,  where $N_{\text{q}} = M_{\text{Tx}}N+1$ and \textcolor{c}{the stacked vector is $\boldsymbol{p}_{\text{x}}= \left [\boldsymbol{p}_{\text{x}_1}^{T},\boldsymbol{p}_{\text{x}_2}^{T},\cdots,\boldsymbol{p}_{\text{x}_k}^{T},\cdots,\boldsymbol{p}_{\text{x}_{N_{\text{u}}}}^{T} \right ]^{T}$ with $\boldsymbol{p}_{\text{x}_k} \in \mathbb{C}^{N_{\text{q}}}$.}

The signaling rate factor $M_{\text{Tx}} >1$  corresponds to faster-than-Nyquist signaling and is related to the oversampling factor by $MM_{\text{Tx}} = M_{\text{Rx}}$. At the BS ideal digital-to-analog converters (DACs) are considered.  The transmit pulse shaping filter $g_{\text{Tx}}(t)$ in discrete time is represented by the Toeplitz matrix  $\boldsymbol{G}_{\textrm{Tx}}$  \textcolor{c}{with $a_{\textrm{Tx}}=  (T /  M_{\text{Tx}} )^{1/2} $  as the normalization factor to  unit  energy}
\begin{align}
\label{eq:GTx}
\boldsymbol{G}_{\textrm{Tx}}=  a_{\textrm{Tx}} \begin{bmatrix}
\left[ \ \boldsymbol{g}_{\text{Tx}}^T \ \right]  \ 0 \cdots \ \ \ 0  \\
0 \ \left[ \ \boldsymbol{g}_{\text{Tx}}^T \ \right] \ 0 \cdots 0 \\
\ddots  \ddots \ddots    \\
0 \cdots \ \ \  0 \ \left[ \ \boldsymbol{g}_{\text{Tx}}^T  \ \right]   
\end{bmatrix}_{N_{\text{tot}}  \times 3N_{\text{tot}} }
\text{.}
\end{align} 

 At the receiver, a pulse shaping filter and 1-bit analog-to-digital converter, process the signal of the $k$th user. The receive filter $g_{\text{Rx}}(t)$  is represented by the Toeplitz matrix 
\begin{align}
\label{eq:GRx}
\boldsymbol{G}_{\textrm{Rx}}=  a_{\textrm{Rx}} \begin{bmatrix}
\left[ \ \boldsymbol{g}_{\text{Rx}}^T \ \right]  \ 0 \cdots \ \ \ 0  \\
0 \ \left[ \ \boldsymbol{g}_{\text{Rx}}^T \ \right] \ 0 \cdots 0 \\
\ddots  \ddots \ddots    \\
0 \cdots \ \ \  0 \ \left[ \ \boldsymbol{g}_{\text{Rx}}^T  \ \right]   
\end{bmatrix}_{N_{\text{tot}}  \times 3N_{\text{tot}} }
\text{,}
\end{align} 
with
$\boldsymbol{g}_{\text{Rx}} = [ g_{\text{Rx}} (-T  ( N + \frac{1}{M_{\text{Rx}}}  )  ),
g_{\text{Rx}}(-T  ( N + \frac{1}{M_{\text{Rx}}} )  + \frac{T}{M_{\text{Rx}}}  ), \ldots,$
$ g_{\text{Rx}} (T ( N + \frac{1}{M_{\text{Rx}}} )  ) ]^T $ as the coefficients of the ${g}_{\text{Rx}} (t)$ filter and $a_{\textrm{Rx}}=  (T /  M_{\text{Rx}} )^{1/2} $  that corresponds to  unit  energy normalization.
The channel matrix $\boldsymbol{H} \in \mathbb{C}^{N_{\textrm{u}}\times N_{\textrm{t}}}$ describes a frequency flat fading channel. 

The effects of pulse shaping filtering are given by the combined waveform determined by the transmit and receive filters which is described by  $ v (t) = g_{\text{Tx}}(t) * g_{\text{Rx}}(t) $ with size ${N_{\text{tot}} \times N_{\text{tot}}}$. The combined waveform is represented by the Toeplitz matrix $\mathbb{\boldsymbol{V}}$,
 \begin{align}
\label{eq:MatrixV_offset}
  \mathbb{\boldsymbol{V}} = \;
   \begin{bmatrix}
      v\left ( 0 \right ) & v\left ( \frac{T}{M_{\text{Rx}}} \right ) & \cdots &  v\left (T N  \right ) \\
      v\left ( -\frac{T }{M_{\text{Rx}}}  \right ) & v\left ( 0 \right ) & \cdots &  v\left (T \left ( N-\frac{1}{M_{\text{Rx}}} \right ) \right ) \\
      \vdots & \vdots & \ddots & \vdots \\
			v\left (-T N \right ) &  v\left (  T \left ( -N+\frac{1}{M_{\text{Rx}}} \right )  \right ) & \cdots &  v\left (0 \right )
   \end{bmatrix}
\end{align}
The matrix $\boldsymbol{U} \in \mathbb{R}^{N_{\text{tot}} \times N_{\text{q}}}$, describes the $M$-fold upsampling operation which relates different signaling
and sampling rates
\begin{align}
\label{eq:Matrizu}
\boldsymbol{U}_{m,n}=
\begin{cases}
  1,  & \textrm{for} \quad m = M \cdot \left ( n-1 \right )+1\\
  0, &  {\textrm{else.}}
\end{cases}
\end{align}
The received signal is quantized and vectorized such that $\boldsymbol{z} \in \mathbb{C}^{N_{\text{u}}N_{\text{tot}}}$ is obtained as
\begin{equation}
\boldsymbol{z} = Q_{1}\left ( \boldsymbol{y} \right ),
\end{equation}
where
\begin{equation}
\boldsymbol{y} = \left (\boldsymbol{H}\boldsymbol{P}_{\text{sp}} \otimes \boldsymbol{I}_{N_{\text{tot}}}\right )\left (\boldsymbol{I}_{\text{N}_{\text{u}}} \otimes \boldsymbol{V}\boldsymbol{U}\right )\boldsymbol{p}_{\text{x}}  + \left (\boldsymbol{I}_{\text{N}_{\text{u}}} \otimes \boldsymbol{G}_{\text{Rx}}\right )\boldsymbol{n}\text{.}
\end{equation}

The vector  $\boldsymbol{n} \in \mathbb{C}^{N_{\text{u}}3N_{\text{tot}}}$ represents the complex Gaussian noise vector with zero mean and variance $\sigma_{n}^{2}$. 
Considering perfect channel state information, the conventional spatial Zero-Forcing (ZF) precoding matrix \cite{SpencerHaardt_2004} is defined as
\begin{align}
\boldsymbol{P}_{\text{sp}} = c_{\text{zf}}\boldsymbol{P}_{\text{zf}}
\end{align}
where
\begin{align}
\boldsymbol{P}_{\text{zf}} = \boldsymbol{H}^{H}\left ( \boldsymbol{H}\boldsymbol{H}^{H}\right )^{-1}
\end{align}
The scaling factor $c_{\text{zf}}$ is given by
\begin{align}
c_{\text{zf}} =  \sqrt{\left (  N_{\text{u}}/ \mathrm{trace} \left (  \left ( {\boldsymbol{H}}{\boldsymbol{H}}^{H} \right )^{-1} \right )  \right )} \text{.}
\end{align}
Several other precoding structures \cite{siprec,bbprec,gbd,wlbd,cqabd,rsbd,mbthp,rsthp} could be considered in this system.
The next section briefly describes the time-instance zero-crossing modulation \cite{Viveros_2023}.
 
\section{TI ZX Modulation}
\label{sec:TIZX}
This section describes the TI ZX modulation proposed in \cite{Viveros_2023}. The  TI ZX modulation conveys the information into the time instances of zero-crossings and considers the absence of zero-crossings as a valid mapping from bits to sequences. Considering $R = 1+ M_{\text{Rx}}$ unique symbols, each symbol \textcolor{c}{$\boldsymbol{x}_{k,j}$, where $j=1,2,\cdots, N$}  is mapped in a
codeword \textcolor{c}{$\boldsymbol{c}_{s_{k,j}}$} of $M_{\text{Rx}}$  binary samples which convey the information according to the time-instances in which the zero crossing occurs or not within the symbol interval, as related in the $c_\text{map}$ assignment in Table~\ref{tab:cmap}. The desired output sequence $\boldsymbol{c}_{\text{out}_{k}}$ is generated by the concatenation of the mapping sequences of each transmit symbol.

\begin{figure}[t]
\begin{center}
\captionsetup{justification=centering}
\includegraphics[height=3cm, width=8cm]{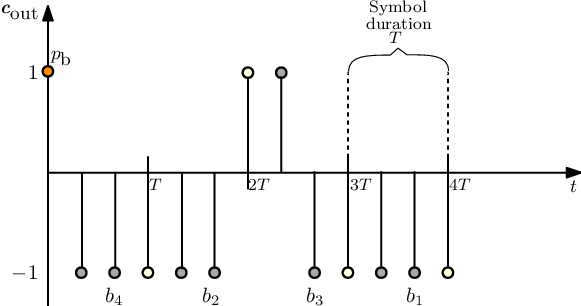}
\caption{Representation of the $\boldsymbol{c}_{\text{out}}$ sequence for $M_{\text{Rx}}=3$.}
\label{fig:ejemploCout}       
\vspace{-0.8em}
\end{center}
\vspace{-0.6em}
\end{figure}

The codewords \textcolor{c}{$\boldsymbol{c}_{s_{k, j}}$} of each transmit symbol need to be concatenated to shape the vector $\boldsymbol{c}_{\text{out}_{k}}$ then, the last sample of the previous segment must be taken into consideration to generate the appropriate codeword that meets the assignments established in $\boldsymbol{c}_{\text{map}}$. In the same way, a pilot sample needs to be added at the beginning of the sequence to map the first transmit symbol.  The last means that each single symbol can be mapped to two different codewords containing the same zero-crossing information. The Gray coding for TIZX modulation, proposed in \cite{Viveros_2023} is also considered for the present study. 
\begin{table}
\caption{Zero-crossing assignment $\boldsymbol{c}_{\text{map}}$}
\label{tab:cmap} 
\normalsize
\begin{center}
\scalebox{0.65}{
\begin{tabular}{|c|l|}
\hline
\multicolumn{2}{|c|}{$\boldsymbol{c}_{\text{map}}$}                                                    \\ \hline
{\color[HTML]{000000} symbol} & \multicolumn{1}{c|}{{Zero-crossing assignment}} \\ \hline
$b_1$                         & No zero-crossing                                                                            \\ \hline
$b_2$                         & Zero-crossing in the $M_{\text{Rx}}$ interval                                                \\ \hline
$b_3$                         & Zero-crossing in the $M_{\text{Rx}}-1$ interval                                              \\ \hline
$\vdots$                      &  \multicolumn{1}{c|}{$\vdots$}                                                                \\ \hline
$b_{R_{\text{in}}-1}$         & Zero-crossing in the second interval                                                         \\ \hline
$b_{R_{\text{in}}}$           & Zero-crossing in the first interval                                                          \\ \hline
\end{tabular}}
\end{center}
\end{table}
The mapping process is done separately and in the same way for the in-phase and quadrature components of the symbols. {Fig.~\ref{fig:ejemploCout} shows an example of the construction of the $\boldsymbol{c}_{\text{out}}$ sequence for $M_{\textrm{Rx}}=3$, when the transmitted symbols are $\boldsymbol{x} = \left [ b_4,b_2,b_3,b_1 \right ] $.}

The detection process is performed as the mapping process, separately and similarly for the in-phase and quadrature components of the received sequence of samples $\boldsymbol{z}_{k}$. The detection is based on the Hamming distance metric \cite{LarsonB_2017}.
The received sequence $\boldsymbol{z}_k$ is segmented into $N$ subsequences \textcolor{c}{$\boldsymbol{z}_{\textrm{b}_j}= [ \rho_{j-1}, \boldsymbol{z}_j ]^T \in \{+1,-1\}^{M_\mathrm{Rx}+1}$, 
where $\rho_{j-1}$} corresponds to the last sample of the previous  \textcolor{c}{$\boldsymbol{z}_{\textrm{b}_{j-1}}$} subsequence. Then the backward mapping process is defined as \textcolor{c}{$\cev{d}: \boldsymbol{z}_{\textrm{b}_j} \rightarrow [\rho_{j-1}, \boldsymbol{c}_{\textrm{s}j}^T]$} \cite{Viveros_2023}. With no distortion each subsequence \textcolor{c}{$\boldsymbol{z}_{\textrm{b}_j}$} has associated a codeword according to $\boldsymbol{c}_{\text{map}}$, and the detection is performed in a simple way. \textcolor{c}{Note that, the subscript $k$ is omitted for simplicity.}

With the presence of noise, invalid segments \textcolor{c}{$\boldsymbol{z}_{\textrm{b}_j}$} may be presented. Therefore the Hamming distance metric is required \cite{LarsonB_2017}  which is defined as
\begin{equation}
    \hat{\boldsymbol{x}}_j = \cev{d}(\boldsymbol{c}), \text{~~with~~} \boldsymbol{c} = \arg \!\!\!\! \min_{\boldsymbol{c}_\mathrm{map} \in \mathcal{M}} \!\! \mathrm{Hamming}(\boldsymbol{z}_{\textrm{b}_j}, \boldsymbol{c}_\mathrm{map}),
\end{equation}
\textcolor{c}{where $\text{Hamming} \; ( \boldsymbol{z}_{\textrm{b}_j},\boldsymbol{c}_{\text{map}}) = \sum_{n=1}^{M_{\text{Rx}}+1}\frac{1}{2}\left | \boldsymbol{z}_{\textrm{b}_{j,n}}- \boldsymbol{c}_{\text{map},n} \right |$ and  $\boldsymbol{c}_\mathrm{map} = [\rho_{j-1},\boldsymbol{c}_{\textrm{s}_j}]^T$, and $\mathcal{M}$ denotes all valid forward mapping codewords.}

The next section describes the temporal precoding optimization based on quality of service constraints.

\section{Optimization problem for QOS temporal precoding}
\label{sec:QOS}
The precoding method proposed in \cite{Viveros_ssp_2021}, minimizes transmit energy ${E_{{0}_k}}$  for a given value of $\gamma$  that corresponds to the minimum distance to the decision threshold. 
The transmitted energy per user can be defined by
\begin{align}
\label{eq_energy}
{\boldsymbol{E}}_{0_k} ={\boldsymbol{p}}_{\text{sp}_{k}}^{\text{H}}{\boldsymbol{p}}_{\text{sp}_{k}}  [ (\boldsymbol{W}{\boldsymbol{p}}_{\text{x}_kI})^{\text{T}}(\boldsymbol{W}{\boldsymbol{p}}_{\text{x}_kI})+
(\boldsymbol{W}{\boldsymbol{p}}_{\text{x}_kQ})^{\text{T}}(\boldsymbol{W}{\boldsymbol{p}}_{\text{x}_kQ})] \notag \text{,}
\end{align}
where  ${\boldsymbol{p}}_{\text{sp}_{k}}$ denotes the $k$-th column of ${\boldsymbol{P}}_{\text{sp}}$ and $ \boldsymbol{W} =  \boldsymbol{G}^{T}_{{\text{Tx}}}\boldsymbol{U}$.

The convex optimization problem is solved separately per user and dimension such that the temporal precoding vector ${\boldsymbol{p}}_{\text{x}_k}$ is obtained.
The convex optimization problem  for a given value $\gamma$ can be expressed as
\begin{equation}
\label{eq:convex1}
\begin{aligned}
& \min_{\boldsymbol{p}_{\textrm{x}_{kI/Q}}}
& &  (\boldsymbol{W}{\boldsymbol{p}}_{\text{x}_{kI/Q}})^{\text{T}}(\boldsymbol{W}{\boldsymbol{p}}_{\text{x}_{kI/Q}})\\
& \text{subject to:}
& & \boldsymbol{B}_{k}\boldsymbol{p}_{\text{x}_{kI/Q}} \preceq  -\gamma \boldsymbol{a} \text{,}
\end{aligned}
\end{equation}
where 
\begin{equation}
\label{eq:convexPar}
\begin{aligned}
& \boldsymbol{B}_{{k}} &=& -\beta \left ( \boldsymbol{C}_{{k}{I/Q}}\boldsymbol{V}\boldsymbol{U} \right)\\
& \boldsymbol{C}_{{k}} &=& \textrm{diag}\left (  \boldsymbol{c}_{\text{out}_{{k}I/Q}}  \right )\\ 
& \boldsymbol{a}  &=& \boldsymbol{1} \in \mathbb{R}^{M_\mathrm{Rx}N+1} \text{.}
\end{aligned}
\end{equation}
The subscript $I/Q$ denotes that the problem is solved separately for the in-phase and quadrature components of the signal. {The constraint in \eqref{eq:convex1} in terms of \textcolor{c}{$\boldsymbol{B}_{{k}}$}, ensures that the received signal after quantization is equal to $\boldsymbol{c}_{\mathrm{out}_{kI/Q}}$ in a noise-free case  and $\beta$ refers to the real-valued beamforming gain}. The symbol $\preceq$ in \eqref{eq:convex1} constrains each element of the vector $\boldsymbol{B}_{k}\boldsymbol{p}_{\text{x}_{kI/Q}}$ to be less than or equal to $-\gamma$ such that the minimum distance of the samples of the received signal to the decision threshold is equal to $\gamma$.
Implicitly, the optimization problem shapes the waveform $y(t)$ at the receiver, which is described in the discrete model by $\boldsymbol{H}\boldsymbol{P}_{\text{sp}}\boldsymbol{V} \boldsymbol{U} \boldsymbol{p}_{\textrm{x}_k}$ for the noiseless case.

Considering the spatial ZF precoder, the total transmit energy ${E_{\textrm{Tx}}}$ can be computed as
\begin{align}
\label{Etx}
{E_{\textrm{Tx}}} = \textrm{trace}\left (   \boldsymbol{P}_{\text{sp}}\boldsymbol{R}_{\textrm{x}_{\textrm{Tx}}}\boldsymbol{R}_{\textrm{x}_{\textrm{Tx}}}^{\textrm{H}}\boldsymbol{P}_{\text{sp}}^{\textrm{H}}  \right ),
\end{align}
where  
\begin{align}
\boldsymbol{R}_{\textrm{x}_{\textrm{Tx}}}  = \left [ (\boldsymbol{G}^{\text{T}}_{{\text{Tx}}}\boldsymbol{U}\boldsymbol{p}_{\textrm{x}_{1}})^{\text{T}};(\boldsymbol{G}^{\text{T}}_{{\text{Tx}}}\boldsymbol{U}\boldsymbol{p}_{\textrm{x}_{2}})^{\text{T}}; \cdots ; (\boldsymbol{G}^{\text{T}}_{{\text{Tx}}}\boldsymbol{U}\boldsymbol{p}_{\textrm{x}_{N_{\text{u}}}})^{\text{T}} \right ].
\end{align}
The next section describes the process to obtain the semi-analytical SER.
\section{{QOS Precoding Performance Bound}}
\label{sec:bound}
This section presents the semi-analytical symbol error rate upper bound for the QOS precoding method \cite{Viveros_ssp_2021} with quality of service constraint regarding the minimum distance to the decision threshold $\gamma$. 
Considering $M_{\text{Rx}}= 3$, $4$ different symbols $b_1,b_2,b_3,b_4$ can be transmitted. For $M_{\text{Rx}}= 2$, sequences of symbols are considered such that $8$ different symbols $b_1,b_2,b_3,b_4,b_5,b_6,b_7,b_8$ can be transmitted. The SER is defined as
\begin{equation}
    \text{SER} = \text{P}_{\text{error}}.
\end{equation}
Considering the probability of correct detection as $\text{P}_{\text{cd}}$, the $\text{P}_{\text{error}}$ probability is defined as  $\text{P}_{\text{error}}= (1-\text{P}_{\text{cd}})$, 
\textcolor{c}{
\begin{align}
    \text{SER} &= 1-\text{P}_{\text{cd}}\\
    &=1- {\text{P}\left ( b \right )}\left (  \sum_{l=1}^{m}\text{P}\left ( \hat{x}_{l}= {b}_{l} |x_{l} ={b}_{l} \right )\right ), 
\end{align}}
where $m =8$ and $m =4$ for $M_{\text{Rx}}= 2$ and $M_{\text{Rx}}= 3$, respectively. $\text{P}(b) = 1/4$ for $M_{\text{Rx}}= 3$ and $\text{P}(b) = 1/8$ for $M_{\text{Rx}}= 2$, since all input symbols have equal probability. Considering the worse case, that all $N_{\text{tot}}$ samples of the temporal precoding vector $\boldsymbol{p}_{x}$ are equal to a value $\gamma$, where $\gamma$ corresponds to the minimum distance to the decision threshold, the probability of correct detection $\text{P}$, can be lower bounded with
\textcolor{c}{
\begin{align}
    \text{P}'\left ( \hat{x}_{l}= {b}_{l} |x_{l} ={b}_{l} \right ) \leq \text{P}\left ( \hat{x}_{l}= {b}_{l} |x_{l} ={b}_{l} \right ).
\end{align}}
With this, the SER upper bound is defined as
\textcolor{c}{
\begin{align}
    \text{SER}_{\text{ub}} =1- {\text{P}\left ( b \right )}\left (  \sum_{l=1}^{m}\text{P}'\left ( \hat{x}_{l}= {b}_{l} |x_{l} ={b}_{l} \right )\right ), 
\end{align}}
The probability density function of the $m$-dimensional multivariate normal distribution is
\begin{align}
\label{mvncdf}
f\left ( \boldsymbol{y},\boldsymbol{\mu},\boldsymbol{\Sigma } \right ) = \frac{1}{\sqrt{\left|\Sigma\right |\left ( 2\pi \right )^m}}\text{exp}\left (-\frac{1}{2}(\boldsymbol{y}-\boldsymbol{\mu})\boldsymbol{\Sigma }^{-1}(\boldsymbol{y}-\boldsymbol{\mu})^{T} \right )\text{,}
\end{align}
where $\boldsymbol{\mu}$ corresponds to the mean vector and $\boldsymbol{\Sigma}$ to the covariance matrix defined as $\boldsymbol{\Sigma} = \text{E}\left \{(\boldsymbol{G}_{\textrm{Rx}}\boldsymbol{n})(\boldsymbol{G}_{\textrm{Rx}}\boldsymbol{n})^{T} \right \}$. Considering the received vector $\boldsymbol{y}_{k}$ before quantization associated with the input vector $\boldsymbol{x}_{k}$, the correct detection probability is defined as
\begin{align}
\text{P}'\left ( \hat{x}_{j}= {b}_{j} |x_{j} ={b}_{j} \right ) =  \int  \cdots \int_{\mathcal{R}}   f\left ( \boldsymbol{y},\boldsymbol{\mu},\boldsymbol{\Sigma } \right ) dy_{1} \cdots dy_{m-1}dy_{m}
\end{align}
The integration regions $\mathcal{R}$ and $\boldsymbol{\mu}$ for each symbol $b_{l}$ are presented in Table~\ref{tab:analyMMDDT3} for $M_{\text{Rx}}= 3$ and in Table~\ref{tab:analyMMDDT2} for $M_{\text{Rx}}= 2$.
\begin{table}[H]
\caption{Integration regions $\mathcal{R}$ for each symbol $b_l$ with $M_{\text{Rx}}= 3$.} 
\label{tab:analyMMDDT3}  
\begin{center}
\resizebox{0.45\textwidth}{!}{
\begin{tabular}{|c|c|c|cc|}
\hline
\multirow{2}{*}{Symbol} & \multirow{2}{*}{$\boldsymbol{\mu}$}                                                 & \multirow{2}{*}{\begin{tabular}[c]{@{}c@{}}Received sequence\\ $\boldsymbol{z}_{i}$ detected as $b_i$\end{tabular}} & \multicolumn{2}{c|}{$\mathcal{R}$}                                                                                     \\ \cline{4-5} 
                        &                                                                        &                                                                                                                     & \multicolumn{1}{c|}{$x_{l}$}                                      & $x_{u}$                                            \\ \hline
\multirow{3}{*}{$b_1$}  & \multirow{3}{*}{$\left [  \gamma, \; \; \gamma, \; \; \gamma, \; \; \gamma \right ]$}    & $\left [ \; \;1, \; \; 1, \; \; 1, \; \; 1 \right ]$                                                                  & \multicolumn{1}{c|}{$\left [\; \; \;0,\; \;\;\;0,\; \; \;\;0,\; \;\;\;0 \right ]$}                   & $\left [  \infty, \infty, \infty, \infty \right ]$ \\ \cline{3-5} 
                        &                                                                        & $\left [ \; \;1, \; \;1, -1, \; \;1 \right ]$                                                                 & \multicolumn{1}{c|}{$\left [\; \; \;0,\; \;\;\;0, -\infty, \; \; \; \;0 \right ]$}          & $\left [ \infty, \infty, \; \; 0, \infty \right ]$      \\ \cline{3-5} 
                        &                                                                        & $\left [  \; \;1, -1, \; \;1, \; \;1 \right ]$                                                                 & \multicolumn{1}{c|}{$\left [\; \;\;0,-\infty, \; \; \; 0, \; \; \; \;0 \right ]$}           & $\left [  \infty, \;\;0, \infty , \infty \right ]$      \\ \hline
\multirow{2}{*}{$b_2$}  & \multirow{2}{*}{$\left [  \gamma, \; \; \gamma, \; \; \gamma,  -\gamma \right ]$}   & $\left [  \; \;1, \; \; 1, \; \; 1, -1 \right ]$                                                                 & \multicolumn{1}{c|}{$\left [\; \; \;0,\; \;\;0,\; \;\;\;0,-\infty \right ]$}             & $\left [  \infty, \infty, \infty, \; \; 0 \right ]$      \\ \cline{3-5} 
                        &                                                                        & $\left [ \; \; 1,  -1, \; \; 1,  -1 \right ]$                                                                & \multicolumn{1}{c|}{$\left [ \; \; \;0,-\infty,\; \; \;0,-\infty \right ]$}       & $\left [  \infty,\;\;0, \infty , \;\;0 \right ]$       
                                                                      \\ \hline
\multirow{1}{*}{$b_3$}  & \multirow{1}{*}{$\left [  \gamma, \; \; \gamma, -\gamma, -\gamma \right ]$}  & $\left [\; \; 1, \; \;1, -1, -1 \right ]$                                                                & \multicolumn{1}{c|}{$\left [\; \; \;0,\; \; \;0,-\infty,- \infty \right ]$}       & $\left [  \infty, \infty, \;\;0, \;\;0 \right ]$           \\ \hline
\multirow{2}{*}{$b_4$}  & \multirow{2}{*}{$\left [\gamma, -\gamma, -\gamma, -\gamma \right ]$} & $\left [\; \;1, -1, -1, -1 \right ]$                                                               & \multicolumn{1}{c|}{$\left [ \; \; \;0,-\infty,-\infty,-\infty \right ]$} & $\left [\infty, \; \;0 \; \; \;0, \; \;0 \right ]$                 \\ \cline{3-5} 
                        &                                                                        & $\left [\; \;1, -1, -1, \; \;1 \right ]$                                                                & \multicolumn{1}{c|}{$\left [\; \; \;0,-\infty,-\infty,\; \;\;\;0 \right ]$}       & $\left [  \infty, \; \;0 \; \; \;0, \infty \right ]$            \\ \hline
\end{tabular}}
\end{center}
\end{table}
\begin{table*}[t]
\caption{Integration regions $\mathcal{R}$ for each symbol $b_l$ with $M_{\text{Rx}}= 2$.} 
\label{tab:analyMMDDT2}  
\begin{center}
\resizebox{0.7\textwidth}{!}{
\begin{tabular}{|c|c|c|cc|}
\hline
\multirow{2}{*}{Symbol} & \multirow{2}{*}{$\boldsymbol{\mu}$}                                                 & \multirow{2}{*}{\begin{tabular}[c]{@{}c@{}}Received sequence\\ $\boldsymbol{z}_{i}$ detected as $b_i$\end{tabular}} & \multicolumn{2}{c|}{$\mathcal{R}$}                                                                                     \\ \cline{4-5} 
                        &                                                                        &                                                                                                                     & \multicolumn{1}{c|}{$x_{l}$}                                      & $x_{u}$                                            \\ \hline
\multirow{4}{*}{$b_1$}  & \multirow{4}{*}{$\left [ \gamma, \; \; \gamma, \; \; \gamma, \; \; \gamma, \; \; \gamma \right ]$}    & $\left [ \; \;1,\; \;1, \; \; 1, \; \; 1, \; \; 1 \right ]$                                                                  & \multicolumn{1}{c|}{$\left [\; \; \;0,\; \; \;0,\; \;\;\;0,\; \; \;\;0,\; \;\;\;0 \right ]$}                   & $\left [ \infty, \infty, \infty, \infty, \infty \right ]$ \\ \cline{3-5} 
                        &                                                                        & $\left [ \; \;1,\; \;1, \; \;1, -1, \; \;1 \right ]$                                                                 & \multicolumn{1}{c|}{$\left [\; \; \;0,\; \; \;0,\; \;\;\;0, -\infty, \; \; \; \;0 \right ]$}          & $\left [ \infty, \infty, \infty, \; \; 0, \infty \right ]$      \\ \cline{3-5} 
                        &                                                                        & $\left [ \; \;1,\; \;1, -1, \; \;1, \; \;1 \right ]$                                                                 & \multicolumn{1}{c|}{$\left [\; \;\;0,\; \;\;0, -\infty, \; \;\; \;0, \; \; \; \;0 \right ]$}          & $\left [ \infty, \infty, \; \; 0, \infty,\infty \right ]$      \\ \cline{3-5} 
                        &                                                                        & $\left [  \; \;1, -1, \; \;1, \; \;1, \; \;1 \right ]$                                                                 & \multicolumn{1}{c|}{$\left [\; \; \;0,-\infty, \; \; \;\; 0,\; \; \;\; 0, \; \; \; \;0 \right ]$}           & $\left [  \infty, \;\;0, \infty , \infty, \infty \right ]$      \\ \hline
\multirow{3}{*}{$b_2$}  & \multirow{3}{*}{$\left [ \gamma, \; \; \gamma, \; \; \gamma, \; \; \gamma,  -\gamma \right ]$}   & $\left [  \; \;1,\; \;1, \; \; 1, \; \; 1, -1 \right ]$                                                                 & \multicolumn{1}{c|}{$\left [\; \; \;0,\; \; \;\;\;0,\; \;\;0,\; \;\;\;0,-\infty \right ]$}             & $\left [  \infty, \infty, \infty,\infty, \; \; 0 \right ]$      \\ \cline{3-5} 
&                                                                        & $\left [ \; \;1,\; \;1, -1, \; \;1, -1 \right ]$                                                                 & \multicolumn{1}{c|}{$\left [\; \;\;0,\; \;\;\;0, -\infty, \; \;\; \;0, -\infty \right ]$}          & $\left [ \infty, \infty, \; \; 0, \infty,\infty \right ]$      \\ \cline{3-5} 
                        &                                                                        & $\left [ \; \; 1,  -1, \; \; 1, \; \; 1,  -1 \right ]$                                                                & \multicolumn{1}{c|}{$\left [ \;\; \;0,-\infty,\; \; \;\;0,\; \; \;\;0,-\infty \right ]$}       & $\left [  \infty,\;\;0, \infty,\infty, \;\;0 \right ]$       
                                                                      \\ \hline
\multirow{2}{*}{$b_3$}  & \multirow{1}{*}{$\left [  \gamma, \; \; \gamma, \; \; \gamma, -\gamma, -\gamma \right ]$}  & $\left [\; \; 1, \; \;1,\; \;1, -1, -1 \right ]$                                                                & \multicolumn{1}{c|}{$\left [\; \; \;0,\; \; \;\;0,\; \; \;\;0,-\infty,- \infty \right ]$}       & $\left [  \infty, \infty,\infty, \;\;0, \;\;0 \right ]$  \\ \cline{3-5}          
&                                                                        & $\left [ \; \;1,-1,\; \;1, -1,  -1 \right ]$                                                                 & \multicolumn{1}{c|}{$\left [\; \;\;0, -\infty, \; \;\; \;0, -\infty,-\infty \right ]$}          & $\left [ \infty,  \; \; 0,\infty, \; \; 0,\; \; 0  \right ]$  \\ \hline
\multirow{1}{*}{$b_4$}  & \multirow{1}{*}{$\left [\gamma, \; \; \gamma, -\gamma, -\gamma, -\gamma \right ]$} & $\left [\; \;1,\; \;1, -1, -1, -1 \right ]$                                                               & \multicolumn{1}{c|}{$\left [ \; \; \;0,\; \; \;0,-\infty,-\infty,-\infty \right ]$} & $\left [\infty,\infty, \; \;0 \; \; \;0, \; \;0 \right ]$                        \\ \hline
\multirow{1}{*}{$b_5$}  & \multirow{1}{*}{$\left [\gamma, \; \; \gamma, -\gamma, -\gamma, \; \; \gamma \right ]$} & $\left [\; \;1,\; \;1, -1, -1, \; \;1 \right ]$                                                               & \multicolumn{1}{c|}{$\left [ \; \; \;0,\; \; \;0,-\infty,-\infty,\; \; \;\;0 \right ]$} & $\left [\infty,\infty, \; \;0 \; \; \;0, \infty \right ]$                        \\ \hline 
\multirow{2}{*}{$b_6$}  & \multirow{2}{*}{$\left [\gamma, - \gamma, -\gamma, -\gamma, \; \;\gamma \right ]$} & $\left [\; \;1, -1,-1, -1, \; \;1 \right ]$                                                               & \multicolumn{1}{c|}{$\left [ \; \; \;0,-\infty,-\infty,-\infty,\; \; \;0 \right ]$} & $\left [\infty, \; \;0 \; \; \;0, \; \;0,\;\;\infty \right ]$                 \\ \cline{3-5} 
                        &                                                                        & $\left [\; \;1, -1,\; \;1, -1, \; \;1 \right ]$                                                                & \multicolumn{1}{c|}{$\left [\; \; \;0,-\infty,\; \; \;0,-\infty,\; \;\;\;0 \right ]$}       & $\left [  \infty, \; \;0, \infty, \; \; \;0, \infty \right ]$            \\ \hline  
\multirow{2}{*}{$b_7$}  & \multirow{2}{*}{$\left [\gamma, - \gamma, -\gamma, -\gamma, -\gamma \right ]$} & $\left [\; \;1,-1, -1, -1, -1 \right ]$                                                               & \multicolumn{1}{c|}{$\left [ \; \; \;0,-\infty,-\infty,-\infty,-\infty \right ]$} & $\left [\infty, \; \;0, \; \; \;0, \; \;0 \; \;0 \right ]$                 \\ \cline{3-5} 
                        &                                                                        & $\left [\; \;1, -1,-1, \; \;1 -1 \right ]$                                                                & \multicolumn{1}{c|}{$\left [\; \; \;0,-\infty,\; \; \;0,-\infty,\; \;\;\;0 \right ]$}       & $\left [  \infty, \; \;0,\; \;\infty, \; \; \;0,\infty  \right ]$            \\ \hline
\multirow{1}{*}{$b_8$}  & \multirow{1}{*}{$\left [\gamma, - \gamma, -\gamma, \; \;\gamma, \; \;\gamma \right ]$} & $\left [\; \;1, -1, -1, \; \;1,\; \;1 \right ]$                                                               & \multicolumn{1}{c|}{$\left [ \; \; \;0,-\infty,-\infty,\; \; \;0,\; \; \;0 \right ]$} & $\left [\infty, \; \;0, \; \; \;0, \; \;\infty,\infty \right ]$                          \\ \hline
\end{tabular}}
\end{center}
\end{table*}
Note that, invalid codewords are also detected as the received symbol \textcolor{c}{$b_{l}$,} therefore, also invalid codewords are included in Table~\ref{tab:analyMMDDT3} and Table~\ref{tab:analyMMDDT2}. Moreover, due to symmetry, only the codewords for positive $\rho$ are considered. 

{Since a given SER value is related to $\gamma$, the optimization problem in \eqref{eq:convex1} can be reformulated as}
\begin{equation}
\label{eq:convex1b}
\begin{aligned}
& \min_{\boldsymbol{p}_{\textrm{x}_{kI/Q}}}
& &  (\boldsymbol{W}{\boldsymbol{p}}_{\text{x}_{kI/Q}})^{\text{T}}(\boldsymbol{W}{\boldsymbol{p}}_{\text{x}_{kI/Q}})\\
& \text{subject to:}
& & \boldsymbol{B}_{k}\boldsymbol{p}_{\text{x}_{kI/Q}} \preceq  -\gamma(\text{SER}) \boldsymbol{a} \text{.}
\end{aligned}
\end{equation}
{Considering the Gray coding for TI ZX modulation from \cite{Viveros_2024}. The approximately upper bound probability of error can be presented in terms of the bit error rate (BER) such that}
{
\begin{align}
\label{BER_ub}
    \text{BER}_{\text{ub}} \simeq 
 \frac{\text{SER}_{\text{ub}}}{n_{s}},
\end{align}}
{where $n_{s}$ corresponds to the number of bits per transmit symbol. For $M_{\text{Rx}}=3$, $2$ bits are mapped in one symbol, and for $M_{\text{Rx}}=2$, $3$ bits are mapped in 2 symbols.}

\section{Numerical Results}
\label{sec:num_results}
The comparison between the semi-analytical and numerical SER for the QOS precoding method is presented in terms of uncoded BER. Alternatively, channel codes such as low-density parity-check codes \cite{ldpc,memd} could also be considered. The transmit filter $g_{\text{Tx}}(t)$ is an RC filter and the receive  filter $g_{\text{Rx}}(t)$  is an RRC filter, where the roll-off factors are $\epsilon_{\text{Tx}} = \epsilon_{\text{Rx}} = 0.22$. The bandwidth is defined with $W_{\text{Rx}}=W_{\text{Tx}} =\left ( 1 + \epsilon_{\text{Tx}} \right )/T$. For a fair comparison, the numerical evaluation considers a transmitter and a receiver with one single antenna. 
For numerical evaluation $\sigma_{n}^2=1$. 
\textcolor{c}{Fig.~\ref{fig:SER} compares the numerical SER for the QOS precoding with \eqref{eq:convex1} and the semi-analytical SER upper bound with noise variance $\sigma_{n}^2 = 1$ for $M_{\text{Rx}}=2$ in (a) and $M_{\text{Rx}}=3$ in (b). For numerical evaluation, sequences of $1$ symbol for $M_{\text{Rx}}=3$ and sequences of $2$ symbols for $M_{\text{Rx}}=2$ were considered.}

\begin{figure*}[t]
\begin{center}
%
%
%
\definecolor{mycolor1}{rgb}{0.00000,1.00000,1.00000}%
\definecolor{mycolor2}{rgb}{1.00000,0.00000,1.00000}%
\definecolor{mycolor3}{rgb}{0.83,0.69,0.22}%

\pgfplotsset{every axis label/.append style={font=\scriptsize
},
every tick label/.append style={font=\scriptsize
}
}

\begin{tikzpicture}[font=\scriptsize
] 
\begin{axis}[%
name=IF1,
width=0.7\columnwidth,
height=0.6\columnwidth,
scale only axis,
ymode=log,
xmin=0.1,
xmax=3,
xlabel={$\gamma$},
xmajorgrids,
ymin=0.001,
ymax=1,
ylabel={SER},
ymajorgrids,
legend entries={Analytical SER upper bound,  Numerical SER},
legend style={at={(1,1)},anchor=north east,draw=black,fill=white,legend cell align=left,font=\tiny}
]

\pgfplotsset{
    every axis/.append style={
        extra description/.code={
            \node at (0.5,-0.25) {(a)};
        },
    },
}

\addlegendimage{smooth,color=green,solid, line width=1.1pt, mark=star,
y filter/.code={\pgfmathparse{\pgfmathresult-0}\pgfmathresult}}
\addlegendimage{smooth,color=red, solid, line width=1.1pt, every mark/.append style={solid, fill=red!50}, mark=square*,
y filter/.code={\pgfmathparse{\pgfmathresult-0}\pgfmathresult}}


\addplot+[smooth,color=green,solid, thick, every mark/.append style={solid, fill=gray!20} ,mark=star, mark repeat=5,
y filter/.code={\pgfmathparse{\pgfmathresult-0}\pgfmathresult}]
  table[row sep=crcr]{%
0.1	0.914059972610619\\
0.2	0.884367758714058\\
0.3	0.848639397269203\\
0.4	0.807041898514962\\
0.5	0.760089182798863\\
0.6	0.70860680969606\\
0.7	0.653662312673236\\
0.8	0.596477697105572\\
0.9	0.538333489018331\\
1	0.480495351158519\\
1.1	0.424121288314415\\
1.2	0.370222365308468\\
1.3	0.319610509547916\\
1.4	0.272899129916675\\
1.5	0.230484167864437\\
1.6	0.19256903555371\\
1.7	0.159181615102377\\
1.8	0.1301949392226\\
1.9	0.105379070255812\\
2	0.0844168277314294\\
2.1	0.0669300549988662\\
2.2	0.0525341785955282\\
2.3	0.0408200152467204\\
2.4	0.0314022529988177\\
2.5	0.0239187759983693\\
2.6	0.0180397901390166\\
2.7	0.0134754017510683\\
2.8	0.00996733600634225\\
2.9	0.00729950402503421\\
3	0.00529577397524239\\
3.1	0.00380411034114869\\
3.2	0.00270788422433998\\
3.3	0.00190533396286674\\
3.4	0.00133005108704676\\
3.5	0.000919904205876954\\
3.6	0.000629050216082616\\
3.7	0.000425904383604125\\
3.8	0.000285517037330041\\
3.9	0.000190867077574652\\
4	0.000122958126839712\\
};

\addplot+[smooth,color=red,solid, thick, every mark/.append style={solid, fill=red!20} ,mark=square*, mark repeat=1,
y filter/.code={\pgfmathparse{\pgfmathresult-0}\pgfmathresult}]
  table[row sep=crcr]{%
0.1	0.64885\\
0.6	0.482\\
1.1	0.2543\\
1.6	0.1018\\
2.1	0.03705\\
2.6	0.00945\\
3.1	0.0016\\
3.6	0.0005\\
};

\end{axis}

\begin{axis}[%
name=IF2,
    at={($(IF1.east)+(80,0em)$)},
		anchor= west,
width=0.7\columnwidth,
height=0.6\columnwidth,
scale only axis,
ymode=log,
xmin=0.1,
xmax=3,
xlabel={$\gamma$},
xmajorgrids,
ymin=0.001,
ymax=1,
ylabel={SER},
ymajorgrids,
legend entries={Analytical SER upper bound,  Numerical SER},
legend style={at={(1,1)},anchor=north east,draw=black,fill=white,legend cell align=left,font=\tiny}
]

\pgfplotsset{
    every axis/.append style={
        extra description/.code={
            \node at (0.5,-0.25) {(b)};
        },
    },
}

\addlegendimage{smooth,color=green,solid, line width=1.1pt, mark=star,
y filter/.code={\pgfmathparse{\pgfmathresult-0}\pgfmathresult}}
\addlegendimage{smooth,color=red, solid, line width=1.1pt, every mark/.append style={solid, fill=red!50}, mark=square*,
y filter/.code={\pgfmathparse{\pgfmathresult-0}\pgfmathresult}}



\addplot+[smooth,color=green,solid, thick, every mark/.append style={solid, fill=gray!20} ,mark=star, mark repeat=5,
y filter/.code={\pgfmathparse{\pgfmathresult-0}\pgfmathresult}]
  table[row sep=crcr]{%
0.1	0.837190637637019\\
0.15	0.816045674676419\\
0.2	0.793552300127733\\
0.25	0.769845015502007\\
0.3	0.745059633532138\\
0.35	0.719346804075146\\
0.4	0.692853200805467\\
0.45	0.665726627276584\\
0.5	0.638120511838155\\
0.55	0.610180527753769\\
0.6	0.582046442827285\\
0.65	0.553862369058375\\
0.7	0.525758356912832\\
0.75	0.497860769094447\\
0.8	0.470292794076504\\
0.85	0.443162343476768\\
0.9	0.416576138143665\\
0.95	0.390623278158462\\
1	0.365392996786101\\
1.05	0.34095318983152\\
1.1	0.317370025459143\\
1.15	0.294699250050583\\
1.2	0.272977076292719\\
1.25	0.252239345832429\\
1.3	0.232506154428246\\
1.35	0.213795298659619\\
1.4	0.196112819349511\\
1.45	0.17945200868076\\
1.5	0.163814427996164\\
1.55	0.149170749240109\\
1.6	0.135505431827073\\
1.65	0.122798470329046\\
1.7	0.111017401124715\\
1.75	0.10011557714117\\
1.8	0.0900650009341951\\
1.85	0.080834774094065\\
1.9	0.0723646847454819\\
1.95	0.0646307583729714\\
2	0.0575877637538951\\
2.05	0.0511851761443004\\
2.1	0.0453759418468397\\
2.15	0.0401380848243462\\
2.2	0.0354123875190685\\
2.25	0.0311708820402682\\
2.3	0.0273737837622031\\
2.35	0.0239773037533386\\
2.4	0.0209466382322983\\
2.45	0.0182635294923261\\
2.5	0.0158824169273776\\
2.55	0.013774227432672\\
2.6	0.0119208229208249\\
2.65	0.0102906768010436\\
2.7	0.00886500385693223\\
2.75	0.007612072201861\\
2.8	0.00652244347859632\\
2.85	0.00557475854279166\\
2.9	0.00475588952243333\\
2.95	0.00404461048047178\\
3	0.00343615835664091\\
3.05	0.00290824879898666\\
3.1	0.00245257359919859\\
3.15	0.00206633433561498\\
3.2	0.0017355799545784\\
3.25	0.00145747968617083\\
3.3	0.00121811006638273\\
3.35	0.00101527552525993\\
3.4	0.000845271947194415\\
3.45	0.000702138357577042\\
3.5	0.00058203562709036\\
3.55	0.00048012947195164\\
3.6	0.000396390452479745\\
3.65	0.000325154593580845\\
3.7	0.00026658217308273\\
3.75	0.000218695409535785\\
3.8	0.000176777414972129\\
3.85	0.000144691641818961\\
3.9	0.000117811876426122\\
3.95	9.56731673955602e-05\\
4	7.70041988871206e-05\\
4.05	6.15898770377488e-05\\
4.1	4.95258027044532e-05\\
4.15	4.03158721368202e-05\\
4.2	3.18463127129132e-05\\
4.25	2.82283950336915e-05\\
4.3	2.01990587413992e-05\\
4.35	1.46947832386779e-05\\
4.4	1.25354599527228e-05\\
4.45	9.08071860727944e-06\\
4.5	8.17553231602375e-06\\
4.55	5.76592617018701e-06\\
4.6	3.73283177113048e-06\\
4.65	2.42435731501178e-06\\
4.7	2.53740834965654e-06\\
4.75	1.79890554496964e-06\\
4.8	1.22653332346179e-06\\
4.85	6.94821401747703e-07\\
4.9	5.45006550556337e-07\\
4.95	4.53290517232752e-07\\
5	3.59312774644849e-07\\
5.05	2.62355138724857e-07\\
5.1	1.49769121882137e-07\\
5.15	1.34545721452994e-07\\
5.2	1.01746296277128e-07\\
5.25	6.9395371671277e-08\\
5.3	5.31112768209496e-08\\
5.35	3.77676625529233e-08\\
5.4	2.97377280578104e-08\\
5.45	1.93764237987892e-08\\
5.5	1.44083311948151e-08\\
5.55	1.12291262954756e-08\\
5.6	7.74181840856869e-09\\
5.65	6.03501071338997e-09\\
5.7	4.72466976653152e-09\\
5.75	3.23222404396972e-09\\
5.8	2.41692932334558e-09\\
5.85	1.89135340722402e-09\\
5.9	1.25851196131066e-09\\
5.95	9.50180267800249e-10\\
6	6.79933442881975e-10\\
};

\addplot+[smooth,color=red,solid, thick, every mark/.append style={solid, fill=red!20} ,mark=square*, mark repeat=1,
y filter/.code={\pgfmathparse{\pgfmathresult-0}\pgfmathresult}]
  table[row sep=crcr]{%
0.1	0.7497\\
0.6	0.5097\\
1.1	0.2609\\
1.6	0.1058\\
2.1	0.0341\\
2.6	0.0089\\
3.1	0.0026\\
};

\addplot[smooth,color=black,only marks, every mark/.append style={solid,fill=black!50}, mark=triangle*]
  table[row sep=crcr]{%
	50 2\\
};\label{MC_MTX3}

\end{axis}

\end{tikzpicture}%
\caption{\textcolor{c}{Semi-analytical and numerical SER comparison for the MMDDT precoding method. In (a) $M_{\text{Rx}}= M_{\text{Tx}}= 2$, $N=2$ and $\sigma^2 = 1$. In (b) with $M_{\text{Rx}}= M_{\text{Tx}}= 3$, $N=1$ and $\sigma^2 = 1$.}} 
\label{fig:SER}    
\end{center}
\end{figure*}

\begin{figure*}[t]
\begin{center}
\input{figures/figure_BER}
\caption{\textcolor{c}{Semi-analytical and numerical BER comparison for the MMDDT precoding method. In (a) $M_{\text{Rx}}= M_{\text{Tx}}= 2$, $N=2$ and $\sigma^2 = 1$. In (b) with $M_{\text{Rx}}= M_{\text{Tx}}= 3$, $N=1$ and $\sigma^2 = 1$.}} 
\label{fig:BER}    
\end{center}
\end{figure*}

The SER for numerical and semi-analytical results considers the same value of $\gamma$. However, for the semi-analytical method, all the codewords are assumed to be constructed with $\gamma$. In the case of the QOS precoding method, $\gamma$ corresponds to the minimum distance to the decision threshold so samples can be larger than $\gamma$. 
\textcolor{c}{Furthermore, considering \eqref{BER_ub}, the results from Fig.~\ref{fig:SER}  are also presented in terms of BER in Fig.~\ref{fig:BER} for $M_{\text{Rx}}=2$ in (a) and $M_{\text{Rx}}=3$ in (b). Note that, the analytical upper bound on BER in Fig.~\ref{fig:BER} (a) is obtained from the SER by $\text{BER}\approx\frac{2}{3}\text{SER}$, and for Fig.~\ref{fig:BER} (b) $\text{BER}\approx\frac{1}{2}\text{SER}$ due to the Gray coding for TI ZX modulation.}

{Departing from the semi-analytical SER results in Fig.~\ref{fig:SER}  the constraint $\gamma(\text{SER})$ considered in the optimization problem \eqref{eq:convex1b}, is shown in Table~\ref{tab:gammaSERMRX3} for $M_{\text{Rx}}= 3$ and Table~\ref{tab:gammaSERMRX2} for $M_{\text{Rx}}= 2$, considering $\sigma^{2} =1$.}
\begin{table}[H]
\caption{$\gamma(\text{SER})$ for $M_{\text{Rx}}= 3$.} 
\label{tab:gammaSERMRX3}  
\begin{center}
\begin{tabular}{|cc|}
\hline
\multicolumn{2}{|c|}{$\gamma(\text{SER})$} \\ \hline
\multicolumn{1}{|c|}{SER}       & $\gamma$ \\ \hline
\multicolumn{1}{|c|}{$10^{-1}$}      & 1.75      \\ \hline
\multicolumn{1}{|c|}{$10^{-2}$}      & 2.65      \\ \hline
\multicolumn{1}{|c|}{$10^{-3}$}      & 3.35        \\ \hline
\multicolumn{1}{|c|}{$10^{-4}$}      & 3.9     \\ \hline
\multicolumn{1}{|c|}{$10^{-5}$}     & 4.45     \\ \hline
\multicolumn{1}{|c|}{$10^{-6}$}    & 4.8         \\ \hline
\end{tabular}
\end{center}
\end{table}

\begin{table}[H]
\caption{$\gamma(\text{SER})$ for $M_{\text{Rx}}= 2$.} 
\label{tab:gammaSERMRX2}  
\begin{center}
\begin{tabular}{|cc|}
\hline
\multicolumn{2}{|c|}{$\gamma(\text{SER})$} \\ \hline
\multicolumn{1}{|c|}{SER}       & $\gamma$ \\ \hline
\multicolumn{1}{|c|}{$10^{-1}$}       & 1.9      \\ \hline
\multicolumn{1}{|c|}{$10^{-2}$}       & 2.75      \\ \hline
\multicolumn{1}{|c|}{$10^{-3}$}       & 3.45     \\ \hline
\multicolumn{1}{|c|}{$10^{-4}$}       & 4     \\ \hline
\multicolumn{1}{|c|}{$10^{-5}$}      & 4.45       \\ \hline
\multicolumn{1}{|c|}{$10^{-6}$}     & 4.9          \\ \hline
\end{tabular}
\end{center}
\end{table}

\section{Conclusions}
\label{sec:conclusiones}
{This work presents a novel precoding technique based on semi-analytical results regarding the \textcolor{c}{SER}. The precoding method is based on a QOS constraint with time instance zero-crossing modulation, where the constraint is defined in terms of a desirable \textcolor{c}{SER}}. The considered system is a bandlimited multi-user MIMO downlink scenario, with 1-bit quantization and oversampling at the receiver. The QOS precoding method minimizes the transmitted energy while taking into account the quality of service constraint.  The semi-analytical SER corresponds to an upper bound since it is considered that all the transmit samples have the same distance to the decision threshold. 
The comparison between numerical and semi-analytical results for $M_\mathrm{Rx} =2$ and $M_\mathrm{Rx} =3$ show that the semi-analytical method corresponds to an upper bound, which is consistent with established theory.



\balance
\bibliographystyle{IEEEbib}
\bibliography{ref}

\end{document}